\documentclass[draft]{agujournal2018}

\draftfalse
\journalname{JGR: Space Physics}

\newcommand{\receive}[1]{\textit{Received {#1}}}
\newcommand{\revise}[1]{\textit{Revised {#1}}}
\newcommand{\publ}[1]{\textit{Accepted {#1}}}

\begin{document}

\title{Investigation of a confined C-class flare in an arch filament system
close to a regular sunspot}
   
\authors{Rohan Eugene Louis}

\affiliation{1}{Udaipur Solar Observatory, Physical Research Laboratory, Dewali 
Badi Road, Udaipur - 313001, Rajasthan, India\\
\jobname .tex compiled on \today\\
\receive{2019 March 01}; \revise{2019 July 23}; \publ{2019 September 23}
}

\correspondingauthor{Rohan Eugene Louis}{rlouis@prl.res.in}

\begin{keypoints}
\item C1.1 class confined flare occurred in an arch filament system close 
to a regular, unipolar sunspot on 2013 September 24
\item Flare driven by small-scale, flux cancellation at neutral line 
underlying arch filament increasing filament twist and destabilizing it
\item Twisted filament expands as it rises producing 2-ribbon flare 
at neutral line but confined by overlying horizontal fields of active region
\end{keypoints}

\justify

\begin{abstract}
A moderate C1.1 class confined flare is investigated here, which occurred on 2013 
September 24 at 22:56~UT, in an arch filament system close to a regular, unipolar 
sunspot.  Spectro-polarimetric observations from the Tenerife Infrared Polarimeter 
at the 70 cm German Vacuum Tower Telescope were combined with data from the Helioseismic 
Magnetic Imager and the Atmospheric Imaging Assembly to identify the processes that 
triggered the flare. The legs of this arch filament were anchored in the leading 
sunspot and the network flux region of opposite polarity. The flare was driven by 
small-scale, flux cancellation at the weak neutral line underlying the arch filament 
which resulted in two small flaring events within an hour of the C1.1 flare. Flux 
cancellation was facilitated by the moat flow from the leading sunspot wherein 
small-scale magnetic fragments stream towards patches of pre-existing flux. The 
cancellation of flux led to the destabilization of the arch filament which was 
seen as an increase in the twist along the arch filament. The horizontal fields 
across the weak neutral line decay faster which cannot prevent the filament 
from rising that results in a two-ribbon flare at the neutral line. The arch 
filament unwinds as it rises, but is confined by the higher, overlying fields 
between the two polarities of the active region that decay much more slowly.
\end{abstract}

\section{Introduction}
\label{intro}
Solar flares are energetic phenomena that result in particle acceleration, 
plasma heating, and the release of energy over the entire electromagnetic 
spectrum. Flares are characterized by a sudden onset and develop on time
scales that are shorter than the photospheric evolution time scale. 
Photospheric motions are responsible for storing magnetic energy in the 
corona which is then impulsively released at a critical point in the evolution 
of the coronal field \citep{2010hssr.book..159F}, resulting in a flare. 
Flares that are associated with coronal mass ejections (CMEs) 
\citep{1992ARA&A..30..113K,1993PhFlB...5.2638G,2000SoPh..194..371P} are termed 
eruptive and tend to occur in regions of a highly non-potential coronal field, 
that overlie the photospheric polarity inversion lines 
\citep{1990ApJS...73..159H,2006ApJ...649..490W}.
While it has been shown that a large majority of X-class flares are associated 
with CMEs \citep{2006ApJ...650L.143Y} and tend to occur in active regions with 
more free energy \citep{2018JGRA..123.1704C}, flares can be non-eruptive or 
confined, failing to produce a CME. A prominent example of confined flares 
are those associated with the largest active region (AR) of solar cycle 24,
NOAA AR 12192. This AR was flare-rich but CME-poor, producing 6 X-class flares, 
22 M-class flares, and 53 C-class, but only 1 CME which resulted from an M4.0 
class flare \citep{2015ApJ...804L..28S,2015ApJ...808L..24C,2016ApJ...828...62J,
2016ApJ...826..119L,2018SoPh..293...16S}. \citet{2018JGRA..123.1704C} analyzed 
58 X-class flares that occurred within 30$^\circ$ of disk center between 1996 
to 2005 and found that 10 X-class flares were confined. 

The source regions of CMEs show a filament channel in the chromosphere 
\citep{2012ApJ...760...81K}, which is a region of strong magnetic 
non-potentiality confined around the polarity inversion line (PIL). 
The filament channel could eventually harbor a filament or prominence, 
wherein the H$\alpha$ fibrils are aligned with the PIL 
\citep{1971SoPh...19...59F,1998SoPh..182..107M}. While a large majority of 
erupting filaments or prominences are closely associated with flares, their 
association with CMEs is about 50\% \citep{2004ApJ...614.1054J,2011MNRAS.414.2803Y}.
Filament eruptions, flares, and CMEs are evidences for the destabilization 
of the coronal magnetic field and their association is still an open ended topic
\citep{2009AdSpR..43..739S}.

Unlike, PIL filaments, arch filament systems (AFS) are a bunch of dark filament threads 
crossing the PIL and connecting the regions of opposite polarity, sometimes extending
into sunspots \citep{1967SoPh....2..451B,1969PASP...81..270W,1985SoPh..100..397Z}.
The arches in the AFS exhibit an upward motion of about 10--15~km~s$^{-1}$ and downwflows 
of about 10--50~km~s$^{-1}$ along their legs. Flares occurring in AFS arise from reconnection
with pre-existing magnetic fields \citep{2008A&A...488.1117Z,2018ApJ...853L..26H} or through
persistent flux cancellation \citep{2018ApJ...855...77S,2018ApJ...859..148W}.
In this article, I investigate a moderate C1.1 class, confined flare, which occurred 
in an AFS outside a regular, isolated sunspot on 2013 September 24. The 
photospheric and coronal conditions are analyzed in order to determine what triggered 
the flare, in the otherwise quiet and small active region, and rendered it confined.

\section{Observations and data analysis}
\label{data}
I carried out spectropolarimetric observations of  NOAA AR 11846 
on 2013 September 24 using the Tenerife Infrared Polarimeter \citep[TIP-II;][]
{2007ASPC..368..611C} at the 70~cm German Vacuum Tower Telescope (VTT), Observatorio 
del Teide, Tenerife, Spain. TIP II provided spectral scans with full-Stokes 
polarimetry in the near infrared region comprising the Fe I line at 10783.05~\AA, 
and the Si I line at 10786.85~\AA. Both lines have a Land\'e $g$ factor of $1.5$ 
and an excitation potential of 3.06~eV and 4.93~eV, respectively 
\citep{2008A&A...488.1085B}.

\begin{figure}[!h]
\centering
\includegraphics[angle=90,width=\columnwidth]{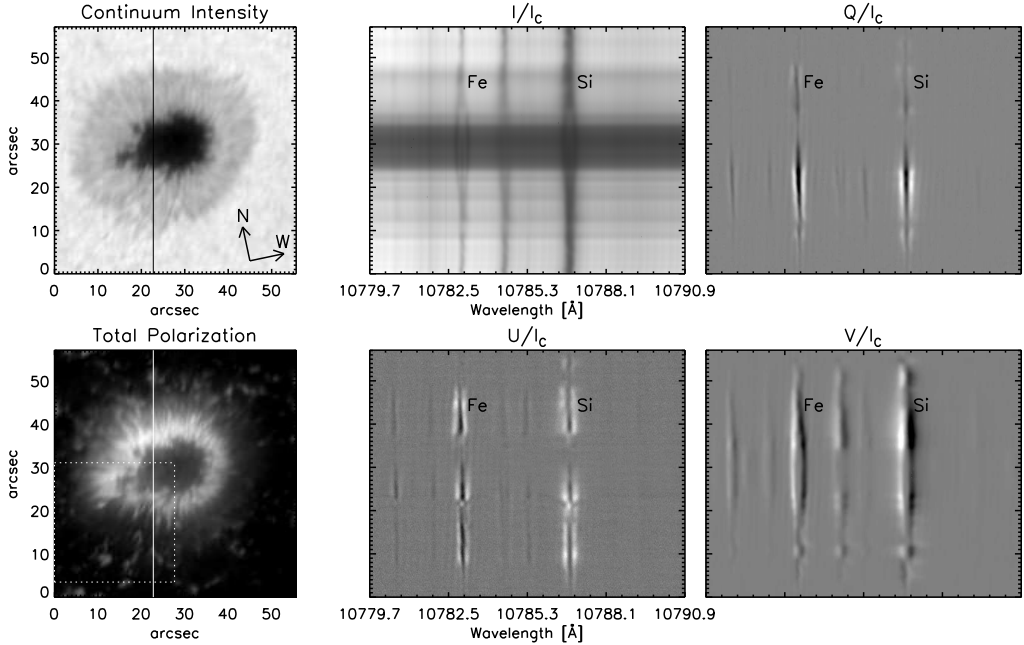}
\vspace{-60pt}
\caption{Near infrared spectropolarimetric observations using TIP-II. The top left panel 
shows the continuum intensity at 10780~\AA. The orientation of the TIP data is indicated 
in the bottom right corner. The bottom left panel shows the map of the total polarization 
derived from the Si line. The white dotted box corresponds to the field-of-view shown in 
Fig.~\ref{fig07}. The middle panels represent the spectral window containing 
the Fe and Si lines, in Stokes $I$ (top) and $U$ (bottom), respectively, while the right 
panels correspond to Stokes $Q$ (top) and $V$ (bottom), respectively. The Stokes spectra 
correspond to the vertical cut shown in the left panels.}
\label{fig01}
\end{figure}

The AR consisted of a unipolar leading sunspot located at at a 
heliocentric angle of 26.6$^\circ$ (143\arcsec, -404\arcsec). At each slit position and 
for a single modulation state, ten exposures of 250 ms each were accumulated. Using 
a 0\farcs35-wide slit, a full scan of the sunspot was carried out with 160 scan steps, 
each step being 0\farcs35. The scan lasted from 9:53--10:26~UT. The spatial sampling 
along the slit was 0\farcs17 while the spectral sampling was 11.05~m\AA. A dichroic 
beam splitter plate, placed in the adaptive optics tank on the first floor of the VTT, 
was used to carry out the scan. A small fraction of light from the beam splitter was 
transmitted to the wave-front sensor (WFS) while the rest of the light was reflected 
to the science focus. A small tilt shifted the image at the science focus but not on 
the WFS. Flat-fielding and polarization calibration were performed at the end of the 
scan, using a rotatable linear polarizer and a phase plate which were introduced in 
the optical path after the exit window of the vacuum tower \citep{1999ASPC..184....3C,
2014AN....335..161L}. The observations were supported by the Kiepenheuer-Institut 
Adaptive Optics System \citep[KAOS;][]{2003SPIE.4853..187V} which provided a highly 
stable scan of the sunspot with a spatial resolution of about 1\arcsec. 
Figure~\ref{fig01} shows the TIP-II observations of the leading sunspot in AR 11846 
along with the Stokes spectra spanning the near infrared region at 1078~nm.

The vector magnetic field and other relevant physical parameters were extracted from 
the TIP-II spectro-polarimetric observations using the Stokes Inversion based on Response 
Functions \citep[SIR;][]{1992ApJ...398..375R} code. SIR computes perturbations in the 
physical quantities at specific locations across the optical depth grid called nodes, 
and then carries out an interpolation to yield values at all grid points. I performed 
a one-component inversion setting the magnetic and dynamic parameters to be constant 
with depth and inverted the Fe and Si lines separately since the lines form in a narrow 
region of the photosphere. This is similar to the strategy adopted by 
\cite{2008A&A...488.1085B}. The temperature stratification was perturbed with five 
nodes. A total of nine parameters were retrieved from the observed profiles, including 
height-independent micro- and macro-turbulent velocities. The vector magnetic field was 
transformed to the local reference frame from the line-of-sight (LOS) frame. The 
180$^\circ$ ambiguity in the azimuth was resolved by assuming a radial variation of 
the horizontal magnetic field and a smooth azimuth \citep{2014A&A...567A..96L}.

I also combined the TIP observations with those from the Helioseismic and 
Magnetic Imager \citep[HMI;][]{2012SoPh..275..229S} as well as the Atmospheric 
Imaging Assembly \citep[AIA;][]{2012SoPh..275...17L}. The HMI data consist 
of continuum intensity filtergrams and line-of-sight (LOS) magnetograms 
having a cadence of 45~s, while the AIA observations comprise filtergrams 
in the He~{\sc ii}~304~\AA~and Fe~{\sc ix}~171~\AA~bands having a cadence 
of 1~min. The HMI and AIA data were selected to span a duration of 3~hr 
starting at 21:00~UT leading up to the C1.1 flare at 22:56~UT. The HMI 
and AIA images were co-aligned using a two-dimensional cross correlation 
routine. AIA images with a cadence of 12~s were also utilized close to the time
of the precursor events as well as the C1.1 flare.

\begin{figure}[!ht]
\centerline{
\includegraphics[angle = 90,width=\textwidth]{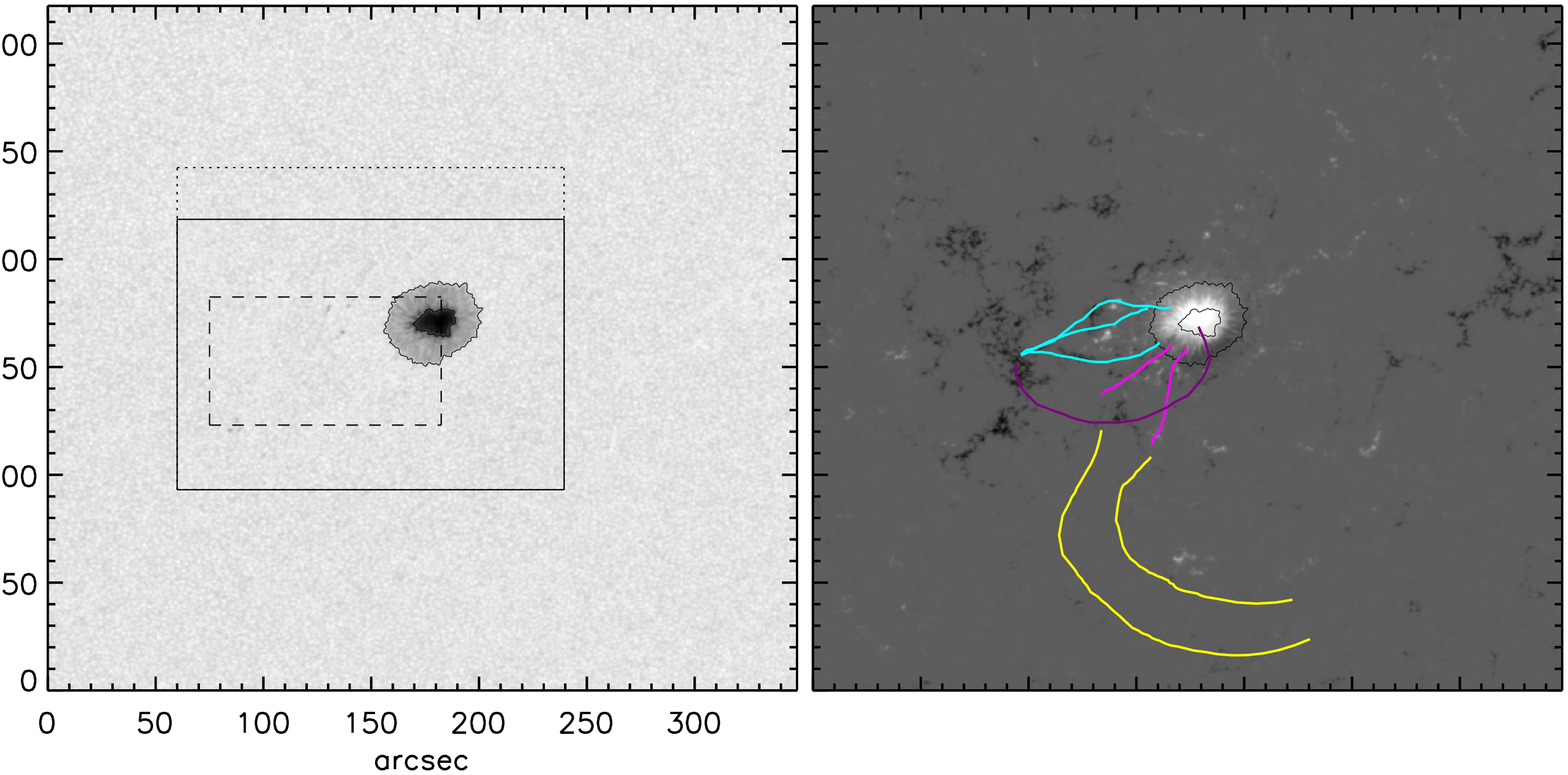}
}
\vspace{-90pt}
\centerline{
\includegraphics[angle = 90,width=\textwidth]{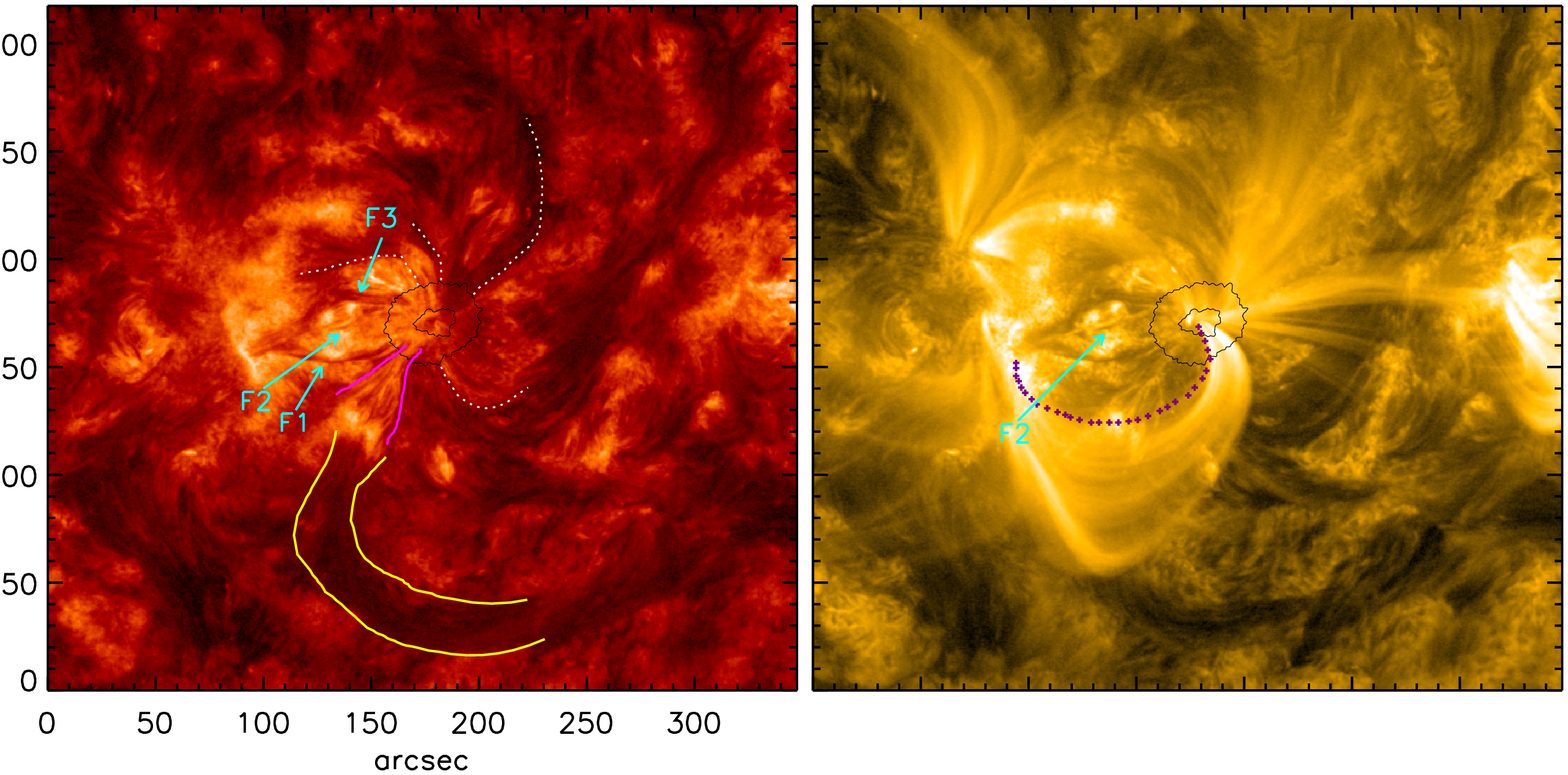}
}
\vspace{-100pt}
\caption{Top panels: HMI continuum image (left) and LOS magnetogram (right) 
of NOAA AR 11846 on 2013 September 24 at 21:00~UT. The {\textit{dashed}}, 
{\textit{dotted}}, and {\textit{solid}} lines in the top left panel correspond 
to the field-of-view shown in Figs.~\ref{fig03}, ~\ref{fig05}, and ~\ref{fig06}, 
respectively. Bottom panels: AIA 
304~\AA~image (left) and 171~\AA~image (right). The white dotted lines
in the lower left panel indicate superpenumbral whirls of the active region with 
a clockwise orientation. The black contours outline the umbra-penumbra and penumbra-quiet 
Sun boundary. North is up and East is to the left. The colours in the bottom 
panels indicate different features (see text) and the same have been overlaid 
on the HMI LOS magnetogram.}
\label{fig02}
\end{figure}

\section{Results}
\label{results}
\subsection{General configuration}
\label{general}
Figure~\ref{fig02} shows NOAA AR 11846 to consist of a single, regular, 
leading sunspot of positive polarity with a diffused network region, about 
90\arcsec~to its east, which is the following polarity of the AR. The 
network region only shows up in the continuum image as a couple of tiny 
pores. The AR appeared close to the Sun's eastern limb on 2013 September 
18 and for a major part of its transit on the disc, remained an $\alpha$-type 
spot. It finally disappeared at the western limb during the early half 
of September 30. During its passage, the AR produced three flares, a C3.9 
class flare at 03:15~UT on September 18, a B4.8 class flare at 21:54~UT on  
September 23, and a C1.1 class flare at 22:56~UT on September 24. 

The lower panels of Fig.~\ref{fig02} show the chromosphere and transition region above 
the AR. The AIA 304~\AA~image (bottom left panel of Fig.~\ref{fig02}) reveals the 
presence of superpenumbral whirls around the AR having a distinct handedness, synonymous 
with a sinistral chirality  
\citep{1998ASPC..150..419M,2000ApJ...540L.115C,2003AdSpR..32.1883M} that is consistent 
with the helicity hemispherical rule. The image also shows three AFS
(marked F1, F2 and F3) all of which originate from the network patch and terminate at/close
to the boundary of the leading sunspot. While F1 and F3 are quite distinct, F2 is extremely 
faint and is more clearly seen as a bright, thread-like structure in the AIA~171~\AA~image 
(bottom right panel). These chromospheric AFS have been marked since the flare occurred in 
this part of the AR, specifically involving F1. For the sake of 
brevity and clarity, I will henceforth refer to the AFS simply as filaments in the text, 
while the term PIL filaments will be used to distinguish the AFS from regular, chromospheric 
filaments.

The projected lengths of the filaments F1, F2, and F3 was estimated to 
be 49, 46, and 56~Mm, respectively. In addition, there exists a large filament channel 
south of the leading sunspot which also exhibits a sinistral chirality (shown in yellow 
in the figure). Hereafter this feature will be referred to as the dark channel since 
there are no clear signatures of an extended PIL filament in 
H$\alpha$(https://gong2.nso.edu/products/scaleView/
view.php?configFile$=$configs/hAlpha.cfg\&productIndex$=$1). The dark channel starts 
from a diffuse patch of negative polarity flux south-east of the sunspot. Short, dark 
fibrils (shown in magenta in the figure) extend from this patch to the outer penumbra 
of the sunspot. The bottom right panel of Fig.~\ref{fig02} shows the structure of the 
AR in the transition region as seen in the AIA 171~\AA~image and indicates several, bright 
loops connecting the leading sunspot to the network patch in addition to the filaments 
seen in the chromosphere. The lowest of these loops is highlighted in purple in the 
figure and is estimated to lie about 24~Mm above F1. The highest of the AIA loops 
connecting the leading sunspot to the network patch are at a projected
height of 100~Mm with respect to F1. Apart from these set of loops, one observes 
connections between the sunspot and opposite polarity flux lying to the north-east and 
west of the former. To the west of the sunspot there is another PIL filament 
that is visible in both AIA bands which has the same handedness as the superpenumbral whirls
associated with the AR. This feature is also seen in H$\alpha$ unlike the dark channel 
described above. The filaments F1, F2, and F3, as well as the overlying loops identified 
in the AIA 304 and 171~\AA~images, respectively, are overlaid on the HMI magnetogram with 
their corresponding colors to provide a context of the magnetic topology of the AR. The 
chromospheric and transition region features identified in Fig.~\ref{fig02} and described above 
correspond to a time close to the occurrence of the flare and nearly 12~hrs after the 
TIP-II observations. However, these features are long-lived, with the exception 
of filament F2, and are seen to be present well before the TIP-II observations, which puts 
their base lifetime at 12~hrs.

\begin{figure}[!ht]
\hspace{25pt}
\includegraphics[angle = 0,width=0.9\textwidth]{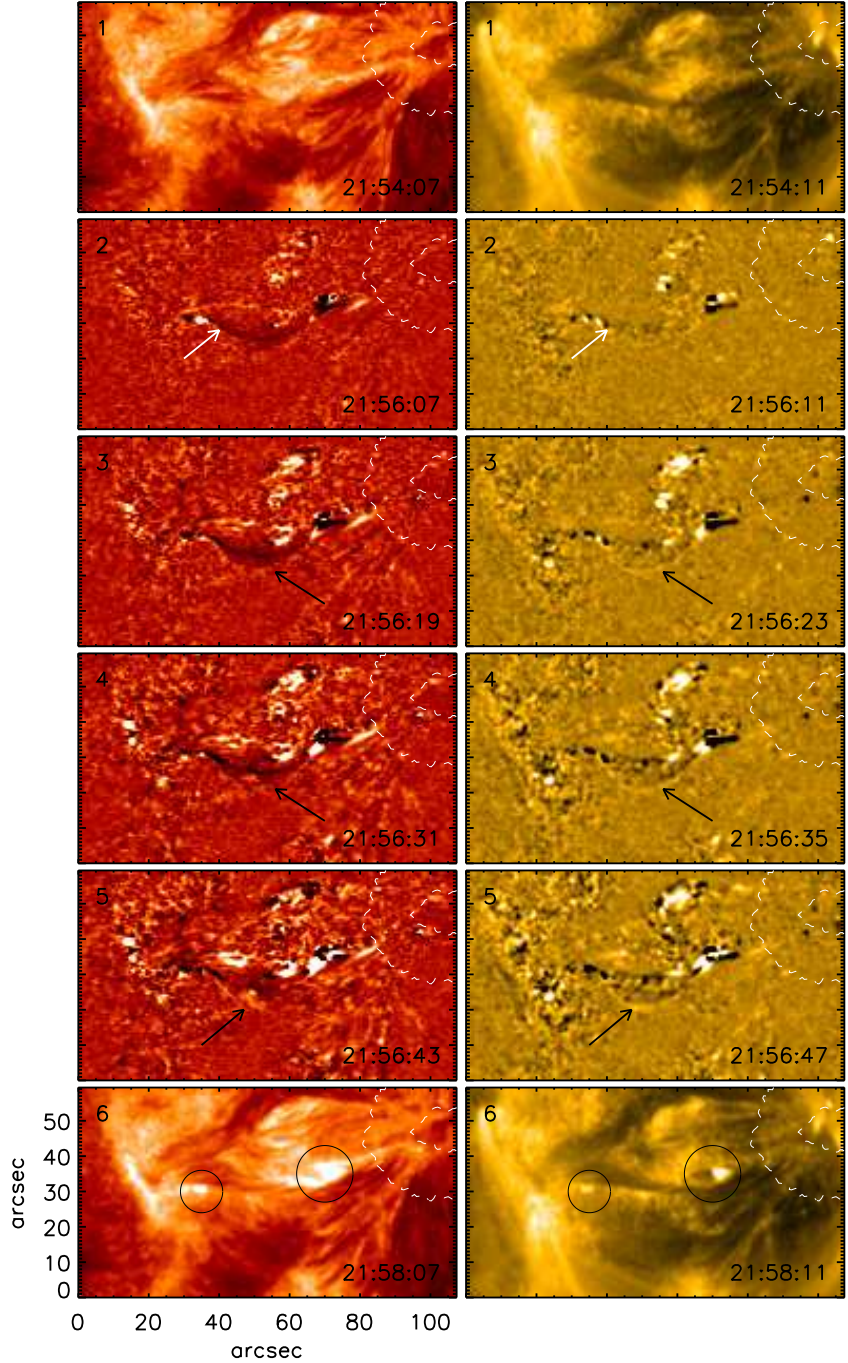}
\vspace{-30pt}
\caption{Time sequence depicting the pre-flaring activity in F1. The 
left and right columns correspond to the AIA 304~\AA~ and 171~\AA~ channels, respectively.
The white and black arrows indicate a set of loops that expand rapidly and detach from the
filament F1, producing small-scale brightenings at the footpoints of the loops enclosed in the 
black circle indicated in panel 6. The images in panels 2 to 5 are base difference images using 
a reference image at 21:55:55~UT. An animation of the figure (movie-01.avi) is available as
supplementary material. }
\label{fig03}
\end{figure}

\begin{figure}[!ht]
\hspace{10pt}
\includegraphics[angle = 90,width=0.85\textwidth]{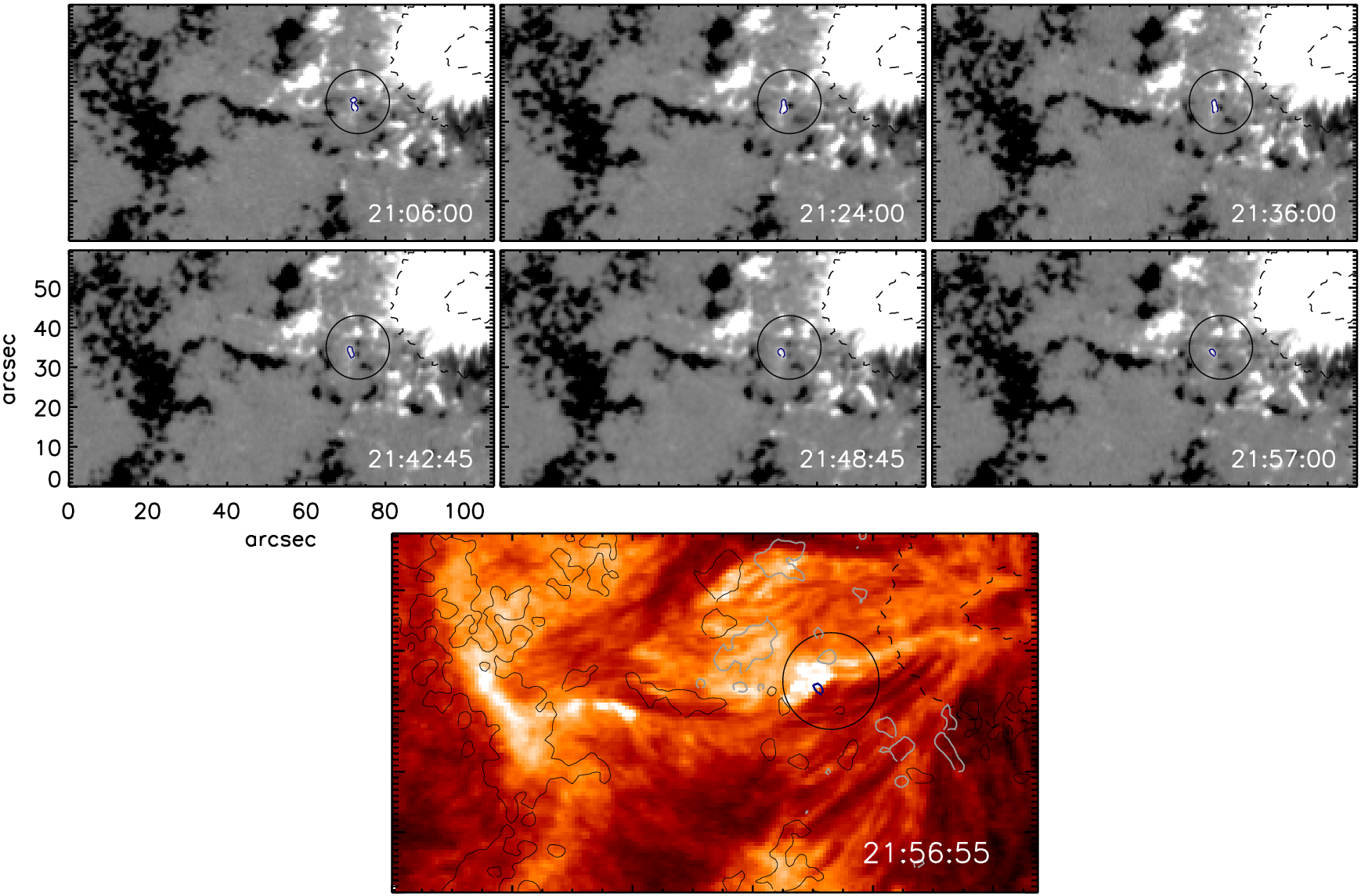}
\vspace{-20pt}
\caption{Flux cancellation event (black circle) leading up to the flaring activity shown 
in Fig.~\ref{fig03}. The dark blue contour outlines a positive polarity patch which indicates
the cancellation of a neighboring magnetic fragment of opposite polarity. The bottom panel shows
an AIA~304~\AA~ image with the one of the footpoint brightenings occurring at the location of flux
cancellation. The black and grey contours correspond to negative and positive polarity, respectively
at a value of $\pm$~100~G. The dashed black lines outline the umbra-penumbra and penumbra-quiet Sun 
boundary of the leading sunspot.}
\label{fig04}
\end{figure}

\subsection{Small-scale flaring activity}
\label{preflare}
In this section, two examples of small-scale flaring activities in the AR, that occur 
within one hour of the C1.1 flare, are described. In addition to these events, the AR 
is also characterized by several other dynamic phenomena, which include brightenings 
along the short dark fibrils (indicated in magenta in Fig.~\ref{fig01}) south east 
of the leading sunspot, and filamentary ejections that start at the location of the 
diffuse polarity south of the leading sunspot and travel southward along the dark 
channel(These can be seen in the animation movie-03.avi that is available 
as supplementary material). Along with the AIA 171~\AA~band, these precursor events are 
also seen in the hotter 193 and 211~\AA~ channels but not in the 335 and 94~\AA~ bands, 
that would suggest that the plasma is heated to at least $10^5$~K. 

\subsubsection{Event 1}
\label{ev1}
Figure~\ref{fig03} shows a sequence of AIA 304~\AA~ and 171~\AA~ images centered on the 
filament F1. Panels 1 and 2 indicate that there is a diffuse, thin, loop-like structure 
which struts out of F1 extending from the eastern end to the central part of F1. The white 
arrow in panel~2 shows the structure much more clearly using a base difference image. 
At 21:55:31~UT, a similar set of loops are seen to unravel from F1, expanding rapidly southwards 
as shown by the black arrow in panels~3--5. This unfurling lasts for about 1~min at the end 
of which remote brightenings are seen at the footpoints indicated by the black circles in 
panel~6. The larger brightening at F1, closer to the leading sunspot, appears at 21:57:31~UT, 
about a 1~min before its counterpart which appears on the southern ridge of F1 close to the 
network flux region. The small flaring activity lasts for about 2.5~min and is analogous 
to the event described in \citet{2015NatCo...6.7008W}. 

The underlying photospheric conditions related to the above flaring event is shown in 
Fig.~\ref{fig04} where there are clear indications of flux cancellation at the location 
of the footpoint brightening. The reduction in flux is more apparently seen in the negative 
polarity fragment that disappears at the interface of the positive polarity fragment outlined 
by the dark blue contour. The decrease in magnetic flux is about 9.8$\times10^{17}$~Mx
and 6.3$\times10^{17}$~Mx in the positive and negative polarity patches, respectively and
occurs over a duration of about 37~min which yields a flux loss rate of 4.5$\times10^{14}$~Mx~s$^{-1}$
2.9$\times10^{14}$~Mx~s$^{-1}$, respectively. The HMI LOS magnetogram has been overlaid on 
the AIA 304~\AA~image for an overall perspective, that shows the filament F1 arching from 
the leading sunspot to the network flux region with a smaller, extended flux patch of negative 
polarity under the axis of F1. The general picture is that F1 overlies a weak PIL with the footpoint
brightenings from the precursor event located on either side of this PIL.

\begin{figure}[!ht]
\hspace{10pt}
\centerline{
\includegraphics[angle = 90,width=0.7\textwidth]{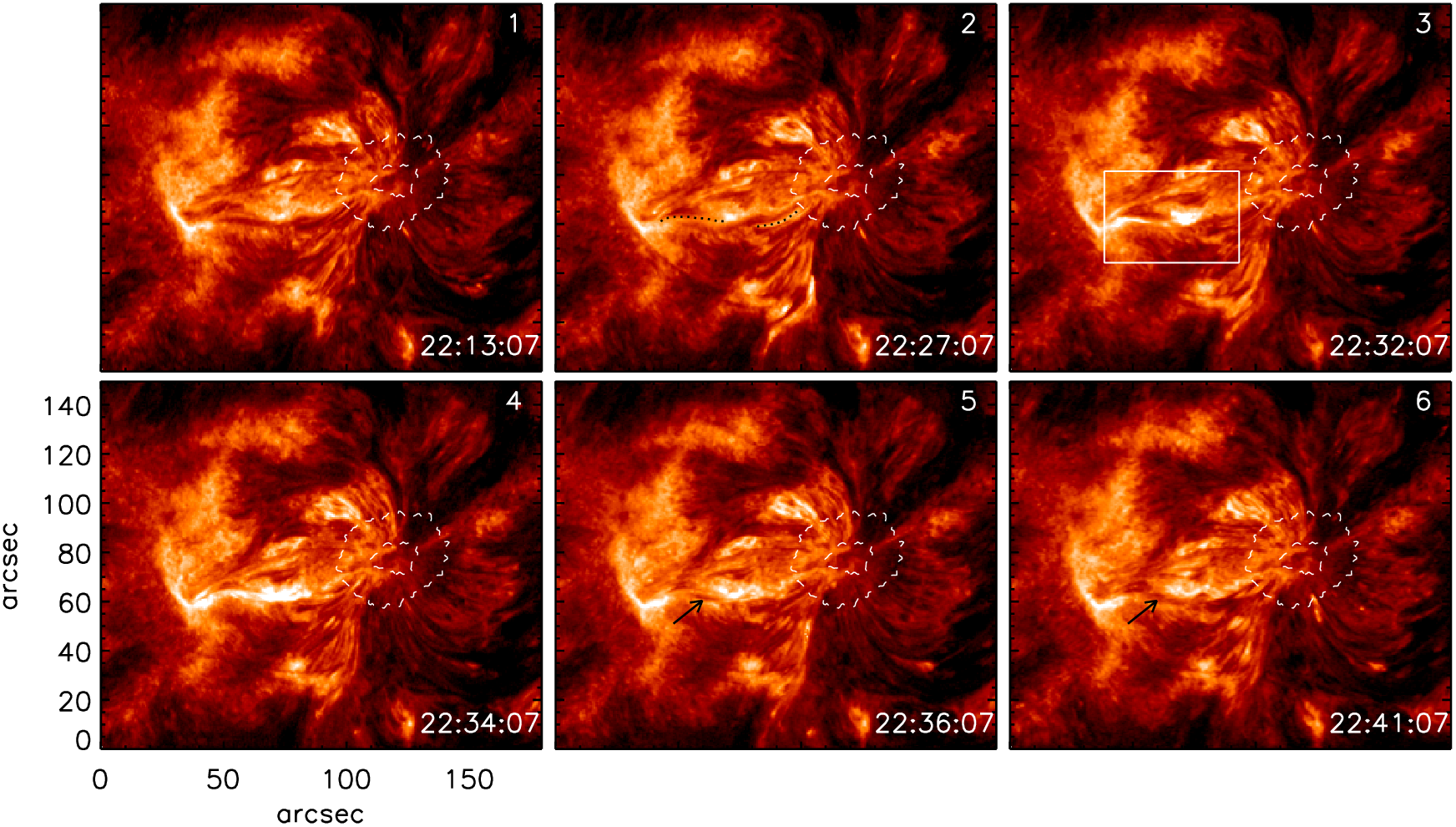}
\includegraphics[angle = 0,width=0.4\textwidth]{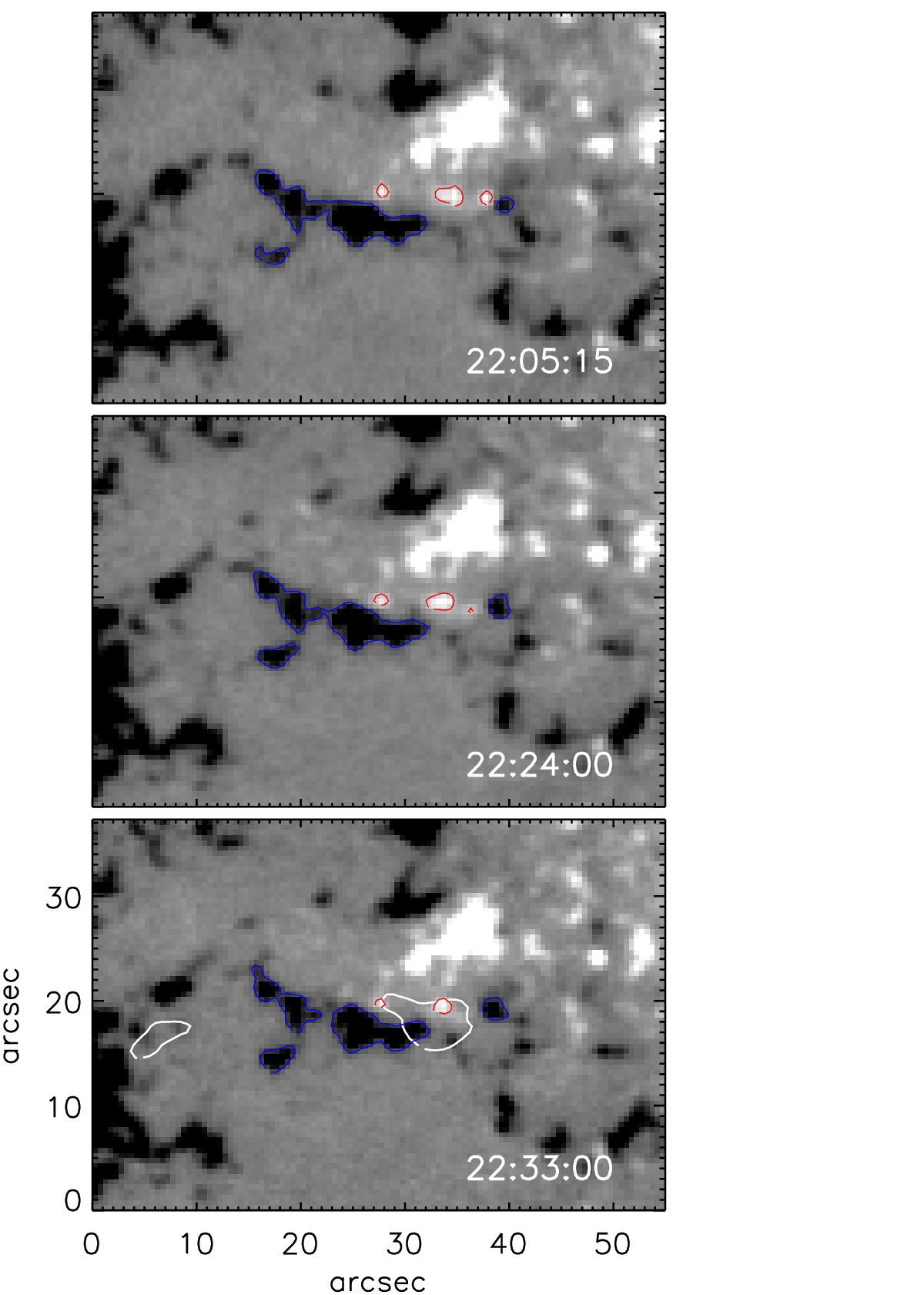}
}
\vspace{-20pt}
\caption{Left: AIA 304~\AA~ images depicting another small flaring event that occurred 
prior to the C1.1 flare. The black dotted line in panel 3 shows a bright thread crossing 
under a dark thread. The white box in panel 3 corresponds to the field-of-view shown on 
the right. The arrows in panels 5 and 6 indicate the injection of cool plasma into the 
filament F1 after the flare. Right: Time sequence of LOS magnetograms showing flux 
cancellation occurring close to the weak PIL under the mid-section of F1 leading to the 
event shown on the left. The blue and red contours correspond to negative and positive 
polarity, respectively at a value of $\pm$~100~G. The white contours in the bottom right 
panel correspond to the footpoint brightenings. An animation of this figure is available 
as supplementary material.}
\label{fig05}
\end{figure}

\subsubsection{Event 2}
\label{ev2}
The second event that was analyzed occurs in nearly the same location as described above 
and is also driven by flux cancellation. The left panels of Figure~\ref{fig05} show a similar 
flaring activity as earlier at 22:33~UT, nearly 30~min after the first event. A comparison 
of panels 1 and 2 in the figure shows that the F1 to exhibit twisted, dark and bright and 
threads (dotted line in panel 2), with the bright thread running under its dark counterpart. 
This intersection point coincides with a small localized brightening which is essentially 
at the weak PIL under F1. The crossing of the bright and dark threads as seen in panel 2, would
suggest that the a twist of least half a turn is present in F1. The stronger brightening 
near the central part of F1 peaks about 
30~sec before its counterpart at the southern ridge of F1 (panels 3 and 4). The flare is 
immediately followed by an injection of cooler plasma along the eastern section of F1 as 
indicated by the black arrows in panels 5 and 6.

The right panel of Fig.~\ref{fig05} shows the time sequence of LOS magnetograms that coincide 
with the second flaring event. The cancellation is primarily seen in the right most patch of 
positive polarity outlined by the red contour. There are only three small fragments of positive 
polarity nearest the larger patch of negative polarity that sits under the central part of F1. 
The flux reduction in these three patches from left to right are 
3.9$\times10^{17}$~Mx, 9.7$\times10^{17}$~Mx, and 8.8$\times10^{17}$~Mx, respectively over a
duration of about 28~min. Using the above values the flux loss rate is estimated to be 
2.4$\times10^{14}$~Mx~s$^{-1}$, 5.8$\times10^{14}$~Mx~s$^{-1}$, and 5.3$\times10^{14}$~Mx~s$^{-1}$,
respectively. While there is a substantial reduction in positive flux, the right most patch
of negative polarity, where the disappearance of the opposite polarity is distinct, exhibits 
an increase of flux amounting to 2.6$\times10^{18}$~Mx. The larger fragment of negative polarity 
on the left outlined by the blue contours, however, shows a reduction in flux of about 
2.4$\times10^{17}$~Mx. The location of the disappearance of the positive flux is about 
10\arcsec~east of the cancellation event described in Sect.~\ref{ev1}. It is also evident 
from the bottom panel that the small-scale changes in photospheric flux are co-spatial with 
the stronger footpoint brightening seen at the intersection of the bright and dark filament 
threads of F1.

\begin{figure}[!ht]
\hspace{10pt}
\includegraphics[angle = 0,width=0.9\textwidth]{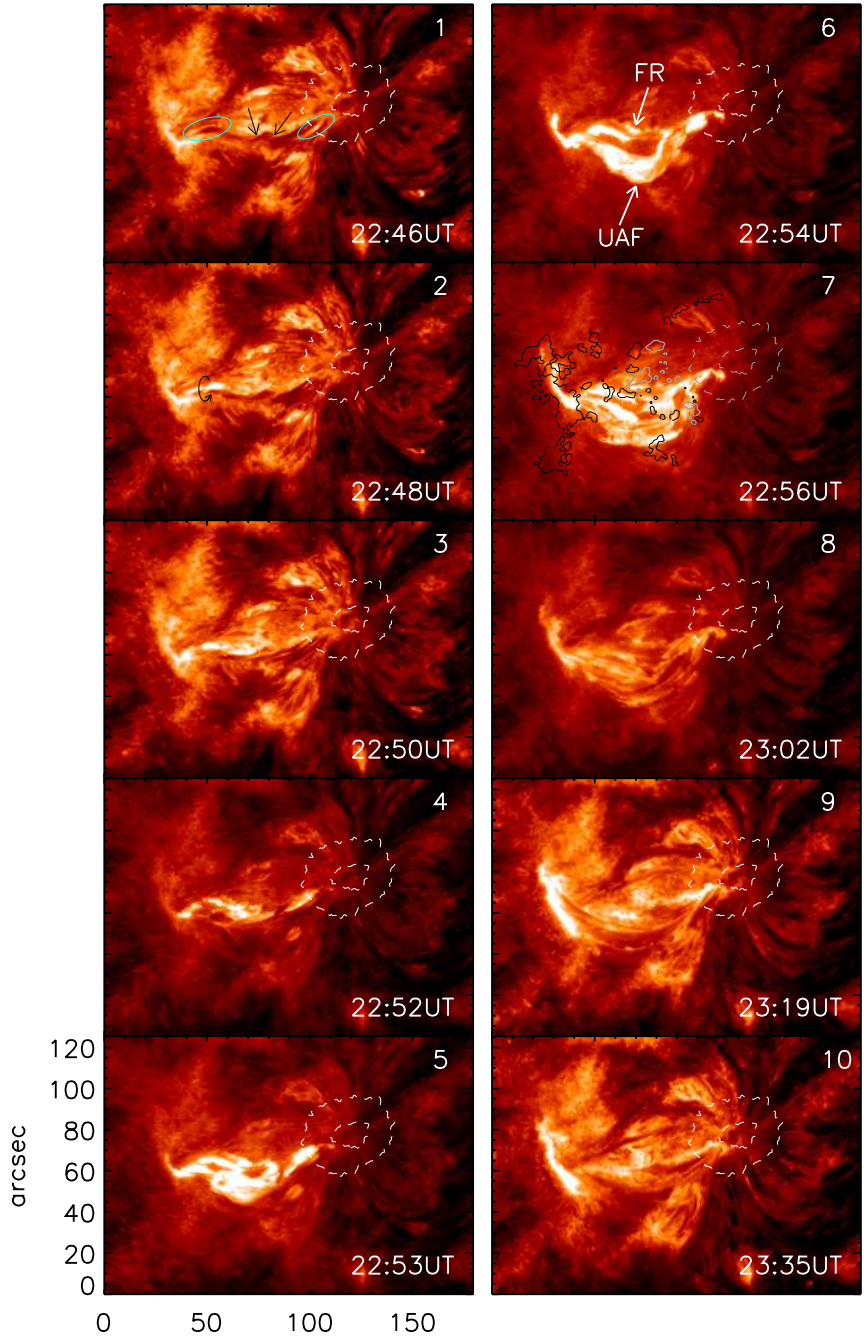}
\vspace{-20pt}
\caption{Evolution of the C1.1 flare seen in AIA~304~\AA~ images. The cyan ellipses in 
panel~1 indicate the onset of the flare with brightenings along threads that twist around 
the axis of F1. The black arrows in the panel indicate an enhancement in the twist in F1. 
Panels~4--8 have been scaled with a higher intensity threshold to show the fine structure 
of the expansion of F1 as it flares. In panel~6, `FR' and `UAF' correspond to flare ribbon 
and untwisting arch filament, respectively. Panel~7 shows contours of the line-of-sight 
magnetic field overlaid on the AIA image with black and grey colors corresponding to 
negative and positive polarity, respectively.}
\label{fig06}
\end{figure}

\subsection{The C1.1 flare in F1}
\label{main}
The precursor flare at 22:33~UT results in the buildup of cool plasma along the filament 
F1 (panel~6 of Fig.~\ref{fig05}). This process continues for about 10~min when filamentary 
brightenings appear at the ends of the filament F1, i.e. at the outer penumbra of the 
leading sunspot and the network flux region of opposite polarity and are indicated by 
cyan ellipses in panel~1 of Fig.~\ref{fig06}. The figure also shows that there is an 
enhancement in the twist along F1 which are depicted by black arrows in the figure and 
can be compared with the filament structure nearly 20~min earlier (panel~2 of Fig.~\ref{fig05}).
The filamentary brightening at the eastern end of F1 intensifies by 22:48~UT which is 
accompanied by an untwisting of F1 at that location in the counterclockwise direction 
as seen along the filament axis from the leading sunspot to the following polarity 
(arrow in panel~2). The unwinding of F1 is accompanied by an increase in the emission 
of the filament that expands southwards as F1 rises (panels~3--5). However, panel~6 
also reveals that along with the rise of F1, the flaring occurs along the weak 
PIL under F1 where small-scale flux cancellation was observed earlier, forming a 
classic two-ribbon structure. The C1.1 flare peaks at 22:55~UT with the flare brightening
extending along F1 to the network flux region in the east, to the umbra-penumbra boundary 
of the leading sunspot. 
The LOS magnetogram contours have been overlaid on the AIA image in panel~7, that show 
the 2-ribbon emission on either side of the neutral line. The flare 
lasts for a duration of 13~min ending at 23:03~UT. The emission associated with the 
flare was located at a distance of about 20~Mm and 23~Mm from the flux weighted 
centroid of the active region, at the beginning and peak time of the event, 
respectively. Panels~9 and 10 of the figure show that the filament structure 
resembles a simple arch with cool plasma suspended between the two polarities of 
the AR. A shorter section of the filament is also reformed at the boundary of the leading 
sunspot near one of the footpoints of the overlying arch. The flare morphology seen in 
the AIA 304~\AA~band is similar in the 171~\AA~channel, with the higher lying set of loops, 
described in Fig.~\ref{fig01}, remaining unaltered during the duration of the flare
and are only weakly displaced southward by about 1~Mm during the expansion of the 
filament.

\begin{figure}[!ht]
\centering
\includegraphics[angle = 90,width=\textwidth]{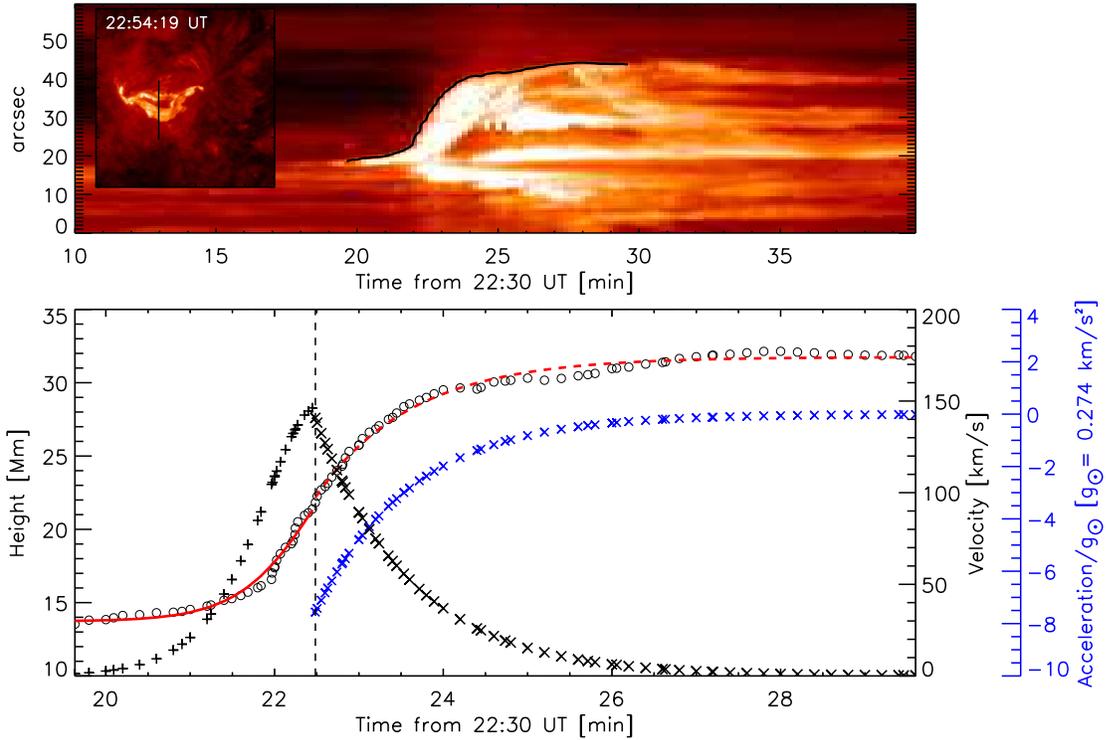}
\vspace{-30pt}
\caption{Kinematics of the rising filament during the flare. Top panel: 
space-time plot along the cut indicated by the black line in the inset at the top 
left corner. The black contour outlines the rising filament and is shown in the bottom 
panel as open circles. The space-time plot has been reversed in the vertical direction. 
Bottom panel: The red lines indicate the fit to the height-time profile during the acceleration 
(continuous red line) and deceleration (red dashed line) phases, respectively,  with the 
values indicated on the left $y-$axis. The corresponding velocity profiles during the two 
phases are shown as \textit{plus} and \textit{cross} symbols, respectively with the scale 
on the right $y-$axis. The deceleration profile is illustrated by blue\textit{crosses} 
whose scale is shown on the right. The vertical dashed line separates the acceleration 
and deceleration phases.}
\label{fig07}
\end{figure}

In order to estimate the speed of the expanding filament during the flare, a space-time
plot was created as shown in Fig.~\ref{fig07}. The inset in the top panel of the figure 
shows the position of the cut with the space-time plot reversed in the vertical direction
for the sake of clarity. The height-time profile was extracted from the black contour 
outlining the top of the erupting filament which attained a maximum height of about 18~Mm 
as indicated in the bottom panel of Fig.~\ref{fig07}. The maximum velocity reached by the 
top of the filament is about 150~km~s$^{-1}$ at about 22:52:29~UT as shown by the 
{\textit{plus}} symbols in the figure. When the flare peaks at 22:55~UT, the filament 
had strongly decelerated and reached 94\% of its maximum height. In comparison, the 
bottom part of the arch filament only rises to a height of about 5.4~Mm.
The velocity in the accelerating phase was determined 
by taking the derivative of a fit to the height-time profile using a Boltzmann-Sigmoid 
function of the form $h_a(t) = a_{0} + (a_{1} - a_{0})/(1 + e^{(a_{2} - t)/a_{3}})$ 
(thick red line). The deceleration phase was fitted with a simple exponential decay curve 
of the form $h_d(t) = b_{0}(1 - b_{1} e^{(-b_{2}t)})$ (dashed red line) and the corresponding 
velocity is shown by {\textit{cross}} symbols. The blue symbols represent the deceleration of 
the filament which is nearly 7.5 times greater than the solar gravitational constant. The 
strong decelerating force prevents the erupting filament from becoming a CME. Only two 
events were reported in the CME catalog provided by CDAW. One was a Halo event that 
occurred at 20:36:05~UT which had a linear speed of 919~km/s and originated from an 
erupting prominence close to the eastern limb (-790\arcsec, 305\arcsec). The other entry 
in the catalog was a very poor event that occurred much earlier at 09:48:05~UT 
and was only seen in C2.

\subsection{Confinement of the C1.1 flare}
\label{confine}

\begin{figure}[!ht]
\centering
\includegraphics[angle = 90,width=0.9\textwidth]{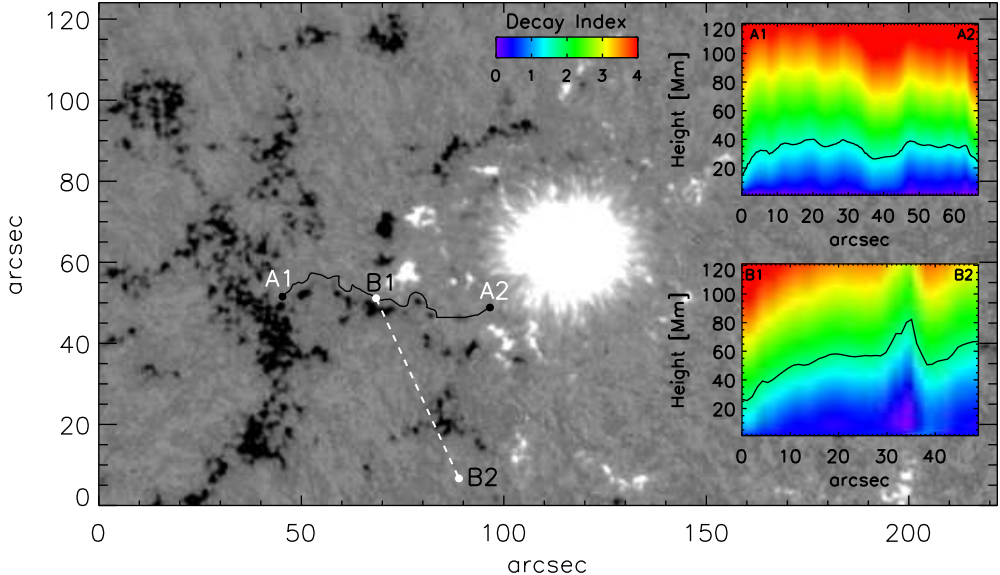}
\vspace{-50pt}
\caption{Confinement of the eruption using the decay index. The background image is 
the vertical component of the magnetic field at 22:36~UT. 
A1--A2 and B1--B2 are two cuts for which the decay index is shown in the insets on the right.
A1--A2 traces the weak PIL under the arch filament, while B1--B2 is along the direction
where the filament expands during the C1.1 flare. The black line in the insets corresponds
to a value of 1.5 for the decay index.}
\label{fig08}
\end{figure}
\begin{figure}[!ht]
\centering
\includegraphics[angle = 90,width=0.9\textwidth]{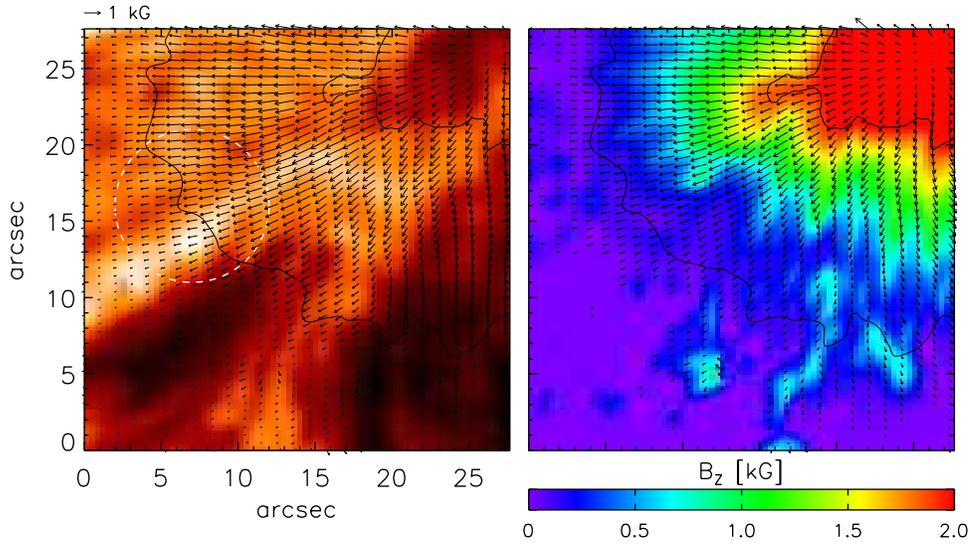}
\vspace{-40pt}
\caption{Vector magnetic field derived from TIP-II observations. The arrows correspond 
to the horizontal magnetic field derived from inversions of the Fe and Si line shown 
in the left and right panels, respectively. The arrows have been drawn for every third 
pixel. The background images correspond to the AIA~304~\AA~image (left) and the vertical 
component of the magnetic field from the Si line inversions (right), whose scaling is 
shown in the color bar below. The dashed white circle in the left panel indicates the 
location of the arch filament F1 anchored to the leading sunspot's boundary.}
\label{fig09}
\end{figure}

I further analyze the conditions that led to the initial eruption of the arch filament and 
its subsequent confinement by computing the decay index from a 
potential field extrapolation \citep{2011ApJ...732...87C,2012ApJ...761...52X,2018ApJ...864..138J}
of the vertical component of the magnetic field prior to the C1.1 flare. The 
decay index is a measure of how fast the horizontal component of the 
overlying strapping fields changes with height. Figure~\ref{fig09} shows the weak PIL 
with A1--A2 denoting its span. The decay index $n$ is calculated over this contour 
as well as along the line B1--B2 which is the direction in which the arch filament 
unwinds and expands. The insets to the right of the figure show $n$ as 
a function of height over the two chosen spatial cuts. For the onset of the torus 
instability \citep{2006PhRvL..96y5002K}, $n$ has a critical value of 1.5, i.e. when the strapping
fields can no longer confine the flux rope. The
black line in the upper inset indicates that the height at which $n = 1.5$
only reaches a maximum value of 40~Mm, and reduces to about 25-30~Mm near the 
central section of the arch filament (near B1).
This indicates that the field decays faster at the PIL and cannot confine the arch filament
from erupting \citep{2015RAA....15.1537J,2018JGRA..123.1704C}. On the other hand, the 
iso-contour of $n = 1.5$ is a lot higher across the cut B1--B2 which 
increases southwards reaching a peak value of 85~Mm which is close to the patch of 
negative polarity about 15\arcsec~north-east of B2. 

It is evident from the above that 
the overlying fields across the weak PIL, which decay faster, are unable to confine 
the arch filament and reconnect as the latter rises, thereby producing the 
two-ribbon flare. However, the second set of overlying fields connecting the leading 
sunspot to the network flux region, decay much more slowly and are able to prevent the 
eruption from developing into a CME. 
%%%Additionally as the event was only a C1.1 flare, 
%%%there may have been insufficient energy to break the overlying fields, which could 
%%%have also rendered the flare confined.

\subsection{Underlying photospheric conditions}
\label{photo}

\subsubsection{Vector magnetic field of the leading sunspot}
\label{sunspot}
In addition to being the strongest and largest concentration of magnetic flux in 
the AR, the leading sunspot also has an important role to play in initiation of 
the flare as the underlying photospheric conditions dictate the overlying 
chromospheric and coronal conditions. This makes it important to determine 
if there exist any signatures in the photospheric magnetic field of the leading 
sunspot that could possibly be related to the filament F1 given that one of its
footpoints is rooted at the boundary of the spot.
Even though the spectropolarimetric observations were acquired nearly 13~hr prior 
to the C1.1 flare, the filament did not exhibit any significant morphological 
changes and thus, a comparison with the photospheric magnetic field is adequate. 
Figure~\ref{fig09} shows the horizontal magnetic field derived from the inversions
of the Fe and Si lines. The left panel shows that the field is predominantly radial
and varies smoothly in the azimuthal direction. The white dashed circle indicates the 
location of F1 at the sunspot boundary (dark filament above a lateral brightening). 
Using the azimuth value at a location on the umbra-penumbra boundary, a line is traced 
outwards till it meets F1 at the sunspot boundary. The azimuth value at this location 
deviated by about 20$^\circ$ from that at the umbra-penumbra boundary. In addition, the 
deviation of the horizontal field from the ideal radial direction as seen from above 
is clockwise, which is consistent with the sign of the twist seen in F1 (Fig.~\ref{fig06}).
However, more detailed modeling is necessary to confirm if this deviation in the 
photosphere is reflected in the structure of the overlying arch filament. The right 
panel of Fig.~\ref{fig09} shows that in the southern quadrant of the sunspot, the 
vertical component of the magnetic field shows disruptions or patches of weaker field
that extend out of the spot's boundary. These patches are related to the dark fibrils
that further connect to the large dark channel seen in the AIA images.

\begin{figure}[!ht]
%%\centering
\hspace{-10pt}
\includegraphics[angle = 90,width=1.0\textwidth]{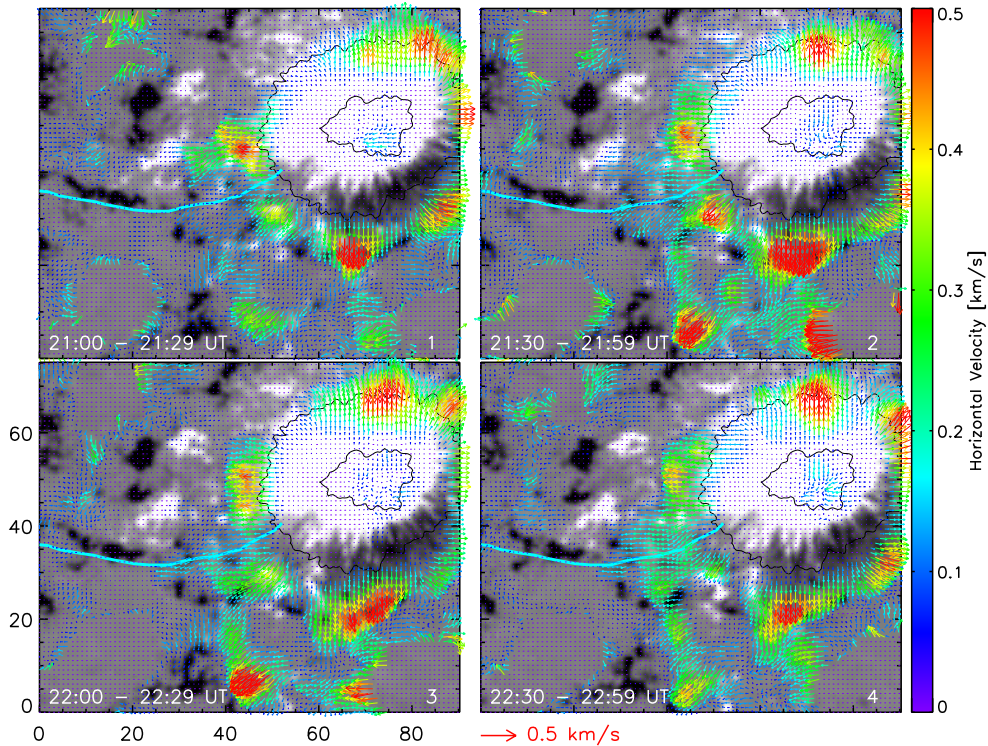}
\vspace{-15pt}
\caption{Horizontal flow maps derived from LCT overlaid on the HMI LOS magnetograms. Each panel
is an average of the horizontal flows over a 30~min time span. The background image is a
LOS magnetogram averaged over the duration indicated in white in the bottom left corner of 
the panel. The thick line in cyan represents the arch filament F1. The $x$- and $y$-axes 
are in units of arcsec. The arrows have been shown for every alternate pixel. The colored 
arrows represent the horizontal speed which are 
indicated in the color bar on the right.}
\label{fig10}
\end{figure}

\subsubsection{Sunspot moat flow driving magnetic flux cancellation}
\label{cancel}
As seen in earlier in Sects.~\ref{ev1} and \ref{ev2}, small-scale flux cancellation 
occurs at the weak PIL directly beneath the filament F1 which drives the precursor 
flares. In this section, the nature of the horizontal flows in the vicinity of F1 
and the leading sunspot are analyzed to determine the manner and extent to which 
flux cancellation is facilitated, and the possible role of the moat flow 
\citep{1972SoPh...25...98S,1988SoPh..115...43B} originating from 
the spot. In order to determine the horizontal flows in the AR, I use local 
correlation tracking \citep[LCT;][]{1986ApOpt..25..392N,1988ApJ...333..427N,2008ASPC..383..373F,
2004ApJ...610.1148W}. LCT computes the relative displacement of small sub-sections in 
an image centered on the pixel of interest. A Gaussian window of a certain
full-width-at-half-maximum (FWHM), that is about the size of the structures that are 
to be tracked, is used to apodize the sub-images. The horizontal speed can thus be 
determined at every pixel given the displacement and the time separation. In order 
to extract the two-dimensional flow, LCT is applied to the HMI LOS magnetogram sequence.
The magnetograms are first smoothed using a $7\times7$ pixel boxcar and then subtracted 
from the original image so that the small-scale magnetic fragments can be tracked easily. 
This scheme is similar to the one adopted by \cite{2004SoPh..225...47R}. An apodizing 
window with a $\sigma=7$~pixels is used for the LCT routine. Although the time 
separation is 45~s, the final horizontal flow field is obtained by averaging 
40 images to retain the persistent motions and suppress the noise.

Figure~\ref{fig10} shows the horizontal flows derived from LCT averaged over a 30~min
interval leading upto the C1.1 flare. Moat flows ranging between 0.3 and 0.5~km~s$^{-1}$ 
are observed outside the sunspot extending to about 10--15\arcsec~ outside the visible 
spot boundary. While the moat flow is nearly uniform around the sunspot's circumference, 
there is a region near the north-east quadrant where the flow is inhibited. This could 
be attributed to the presence of a large, magnetic fragment of positive polarity just 
outside the spot boundary and well within the moat radius, which obstructs the outward 
flow. This is in agreement with the observations of \cite{2008ApJ...679..900V}. The 
moat flow also persists for a period of 2~hrs which reflect the continuous outflow 
from the sunspot that is functional on longer timescales. In general, the figure indicates 
that the horizontal flow is greatly reduced in the neighborhood of pre-existing 
flux that do not exhibit any proper motion, but rather, evolve in-situ. 

The outward flow from the sunspot, particularly in the south-eastern quadrant is also 
directed along the axis representing the filament F1 which is shown in cyan. These 
flows which are about 0.2--0.3~km~s$^{-1}$ in magnitude extend upto a distance of about 
20\arcsec~ into F1. The flux cancellation event associated with the first precursor flare 
is aided by a horizontal flow of about 0.2~km~s$^{-1}$ that is directed from west to east.
The flows reduce considerably once they encounter the large, pre-existing flux patch 
of negative polarity which lies underneath the central part of F1 and is evident from 
panel 2 of the figure. In the case of the second precursor flare, the flux cancellation
occurs 10\arcsec~further east where the flows are considerably weaker till 22:30~UT,
with values of about 0.1~km~s$^{-1}$ (panel 3). However, during the course of the C1.1 
flare, the flows increase to 0.2~km~s$^{-1}$ at this location as seen in panel 4 of the 
figure. The figure also shows that the horizontal flows at the weak PIL (region around 
$x=20$\arcsec, $y=35$\arcsec) are strongly inhibited, which could be attributed to 
magneto-convection and remain unaffected by the spot's moat flow. Horizontal flows of 
about 0.5~km~s$^{-1}$ are also seen to be directed into the dark channel 
(marked in yellow in Fig.~\ref{fig02}) that starts about 20--25\arcsec~ from 
the southern boundary of the sunspot.

\section{Discussion}
\label{discuss}
A C1.1 class, confined flare is analyzed in this article that occurred in an arch filament 
system very close to a regular, unipolar, isolated sunspot. While the continuum images 
do not show any distinguishable presence of a following polarity of the active region, 
the LOS magnetograms reveal a rather diffuse, but extended network flux patch.  
Chromospheric images from AIA indicate that there are three arch filaments, rooted
in the following polarity with the two larger ones extending upto the boundary of the
leading sunspot. In addition to the arch filaments, there are overlying hotter loops 
connecting the two major polarities of the AR. The flare occurred in the southern arch 
filament. Spectro-polarimetric observations from the TIP-II instrument at the 70~cm VTT, 
show a small deviation of the horizontal magnetic field from a radial orientation close
to the sunspot's boundary where the arch filament is oriented. The sense of the deviation
of the magnetic field is consistent with the handedness of the arch filament although detailed
modeling is necessary to substantiate this.
The PIL between the leading sunspot and the network flux region, although devoid of any complex
magnetic structures, nevertheless comprises a few, small, magnetic fragments of mixed polarity. One 
such set of fragments forms a very weak PIL that is located directly beneath the arch filament. 

The C1.1 flare is preceded by two, small-scale flaring events that occur within an hour of 
the former. The first event comprises a set of loops that expand rapidly and detach from 
the arch filament producing brightenings at the loop footpoints on either side of the weak PIL
underlying the arch filament. This precursor flare is similar to the event described in 
\citet{2015NatCo...6.7008W}. The other precursor event, which occurs about 30~min later, 
also produces similar, but stronger, footpoint brightenings at/close to the same location 
as the earlier event. A section of the arch filament is filled by cool plasma immediately 
after the precursor flare that is often seen after an eruption 
\citep{2000ApJ...537..503G,2001ApJ...549.1221G,2007ApJ...661.1260L,2014AN....335..161L}.
Following the precursor flaring events, the arch filament is seen to exhibit an 
increased number of intersections of bright and dark threads suggesting an increase in its twist.
The C1.1 flare involves a fast rise and expansion of the arch filament while also producing 
a two-ribbon emission at the weak PIL. The flare results in the formation of a more relaxed 
arch connecting the two polarities of the active region. The flare emission is located close 
to the centroid of the active region which is in good agreement with the estimates provided by 
\cite{2007ApJ...665.1428W} and \cite{2011ApJ...732...87C} for confined flares. 

The photospheric driver of the precursor events, leading upto the C1.1 flare, is small-scale 
flux cancellation at the weak PIL underlying the arch filament. The flux loss rate is one to 
two orders of magnitude smaller than those associated with surges and EUV jets 
\citep{2004ApJ...610.1136L,2007A&A...469..331J}. Nevertheless, flux cancellation is co-spatial 
as well as co-temporal with the precursor flare brightenings and there are no visible or 
detectable signatures of flux emergence occurring in the region close to the arch filament.
One also finds that the moat flow from the sunspot aids in flux cancellation, particularly 
near the end of the arch filament that is rooted at the boundary of the sunspot. The moat 
flow is responsible for pushing small magnetic elements into the path of pre-existing 
patches of opposite polarity. 

It is known that flux cancellation can lead to the accumulation of helicity 
in the corona \citep{1989SoPh..121..197L,1989ApJ...343..971V,2000SoPh..197...75M,
2000ApJ...529L..49A,2001JGR...10625165L,2001ApJ...548L..99Z,2007PASJ...59S.823S,
2010A&A...521A..49S,2011A&A...526A...2G,2012ApJ...759..105S,2013SoPh..283..429B,
2017NatAs...1E..85W}. Flux cancellation at the PIL under the arch filament produce
the precursor flaring events and introduce more twist in the arch filament, eventually 
destabilizing it \citep{1981SoPh...73..289H,2008ApJ...680.1508L,2013ApJ...779..157G,
2016NatCo...711837X} through the helical kink instability \citep{2004A&A...413L..27T}. 
While an extrapolation of the photospheric magnetic field into the overlying corona 
is necessary to determine the amount of twist in the arch filament following the 
precursor activities, one can qualitatively determine the twist through the number 
of dark and bright windings on the filament as shown in Fig.~\ref{fig06}. One can 
visualize at least two dark threads which would suggest that the number of turns 
in the filament would be $1$ or $2\pi$~radian. This lower bound value of the twist 
is indicative of a weakly twisted, low-lying, flux rope \citep{2015ApJ...809...34C}. 
In an alternate scenario, flux ropes can also be formed during strong, confined flares 
quickly leading upto a major CME within a short duration of time
\citep{2018ApJ...867L...5L}.

The two-ribbon flare is the result of the unstable arch filament rising, causing the 
overlying fields across the weak PIL to reconnect. The overlying fields connecting 
the major polarities of the active region, which decay much more slowly, exert a 
strong tension force in the downward direction which brakes the erupting arch filament,
causing it to decelerate \citep{2005ApJ...630L..97T}. This is evident from 
Fig.~\ref{fig07} where the deceleration exceeds the solar gravitational constant by 
a factor of nearly 8. The height-time profile of the rising filament is similar to the 
findings of \citet{2003ApJ...595L.135J} for a confined filament eruption where the 
filament reached a maximum height of about 80~Mm and a speed of about 200~km~s$^{-1}$. 
The height and speed with which the filament rises is dependent on the strength of the 
axial magnetic field and the twist of the filament 
\citep{2006A&A...460..909M,2008A&A...492L..35A}. Furthermore, being a weak flare of 
C1.1 class, there likely would have been a reduced Lorentz force impulse at the onset 
of eruption to overcome the tension force from the higher overlying fields 
\citep{2018ApJ...858..121L}.

Following the flare, the arch filament relaxes to a simple loop-like structure as seen
in panels 9 and 10 of Fig.~\ref{fig06}. The loss of twist could be attributed to magnetic 
reconnection occurring in either of the following scenarios. Firstly, the filament could 
have underwent internal reconnection, arising from the buckling of the filament axis on the 
western leg that is evident during the peak of the flare (panel 6 of Fig.~\ref{fig06}). 
This distortion of the axis likely brings one part of the filament close to another 
section allowing magnetic reconnection and thus removing the overall twist of the 
filament. The alternate scenario is that, as the filament rises the outermost set 
of twisted field lines reconnect with the lowermost section of the overlying 
fields connecting the two major polarities of the active region. Although the axial field
has the same direction for both sets of field lines, it could be that certain sections of
the filament have a component of the magnetic field where reconnection could be initiated. 
In either of the above cases, one  observes elongated brightenings along the upper part 
of the western half and the lower part of the eastern half of the arch filament (panel 7 
of Fig.~\ref{fig06}) which suggest that reconnection is responsible for the loss of twist 
in the filament rendering a simple potential configuration at the end of the flare.

The active region did produce two additional flares along with the C1.1 flare 
on September 24, which included a C3.9 flare on September 18 and a B4.8 flare on 
September 23. Interestingly, the C3.9 flare was the only one out of the three flares, 
which produced a CME, but the active region was just moving across the eastern limb 
of the Sun at that time. The CME had a linear speed of 208~km~s$^{-1}$ and an angular 
width of 52$^\circ$. However, the active region was still an $\alpha$-type at 
that time and remained so for the entirety of its passage across the disc. Hence, 
it would be interesting to investigate what facilitated the eruption into an ensuing CME 
given that the photospheric magnetic conditions were similar to the event described 
here where the eruption remained confined.

\section{Conclusion}
\label{conclude}
A C1.1 class, confined flare is investigated in this article which occurred on 2013 
September 24 at 22:56~UT, in an arch filament system close to a regular, unipolar 
sunspot. The legs of this arch filament were anchored in the leading sunspot of 
positive polarity and the network flux region of negative polarity. The flare was 
driven by small-scale flux cancellation at the weak PIL underlying the arch filament 
which resulted in two, small flaring events within an hour of the C1.1 flare. Flux 
cancellation was aided by the moat flow from the leading sunspot which was also one 
of the anchorage points of the arch filament. The cancellation of flux led to the 
destabilization of the arch filament which was seen as an increase in the twist 
along the arch filament. The horizontal fields across the weak PIL decay faster 
which cannot prevent the filament from rising which resulted in a two-ribbon flare 
at the PIL. On the other hand, the overlying fields between the two polarities of 
the active region, decay much more slowly and exert a strong downward directed tension 
force which decelerates the rising filament and prevents it from developing into a CME. 
It would be worthwhile to investigate if the above photospheric and coronal conditions
are generic to flares occurring in simple, bipolar active regions. 

\acknowledgments
The Vacuum Tower Telescope is operated by the Kiepenheuer-Institute for Solar Physics 
in Freiburg, Germany, at the Spanish Observatorio del Teide, Tenerife, Canary Islands.
The observing campaign at the Vacuum Tower Telescope was supported by the German 
Science Foundation (DFG) under grant DE 787/3-1.
HMI data is courtesy of NASA/SDO and the HMI science team. They are provided
by the Joint Science Operations Center -- Science Data Processing at Stanford University.
This work was (partly) carried out by using Hinode Flare Catalogue 
(\url{https://hinode.isee.nagoya-u.ac.jp/flare\_catalogue/}), which is maintained by 
ISAS/JAXA and Institute for Space-Earth Environmental Research (ISEE), Nagoya University. 
This work made use of the SOHO LASCO CME Catalog which is generated and maintained at the 
CDAW Data Center by NASA and The Catholic University of America in cooperation with the 
Naval Research Laboratory. SOHO is a project of international cooperation between ESA and 
NASA. REL is grateful to the reviewers for their useful comments which significantly 
improved the article.

\bibliography{ref}

\end{document}